\documentclass[9pt]{article}

\usepackage[compact]{titlesec}

\usepackage[noadjust]{cite}

\usepackage{graphicx}

\usepackage[a4paper,top=1.4cm,bottom=1.4cm,left=1.7cm,right=1.7cm,marginparwidth=1.25cm]{geometry}

\usepackage{caption}

\usepackage{multicol}

\usepackage{subcaption}
\usepackage{float}

\title{\bf Clinical Parameters Prediction for Gait Disorder Recognition \footnote{\textit{all authors} have had equal contribution to the work}}
\author{Soheil Esmaeilzadeh \\
\small Stanford University \\ 
{\tt\small soes@stanford.org}
\and
Ouassim Khebzegga \\
\small Stanford University\\
{\tt\small ouassimk@stanford.edu}
\and
Mehrad Moradshahi \\
\small Stanford University\\
{\tt\small mehrad@stanford.edu}
\footnotetext{footnote with two references} 
}

\date{ }
\usepackage[font=small,skip=0pt]{caption}
\begin{document}

\maketitle

\section{Introduction}
Being able to predict clinical parameters in order to diagnose gait disorders in a patient is of great value in planning treatments. It is known that \textit{decision parameters} such as cadence, step length, and walking speed are critical in the diagnosis of gait disorders in patients. This project aims to predict the decision parameters using two ways and afterwards giving advice on whether a patient needs treatment or not. In one way, we use clinically measured parameters such as Ankle Dorsiflexion, age, walking speed, step length, stride length, weight over height squared (BMI) and etc. to predict the decision parameters. In the second way,  we use videos recorded from patient's walking tests in a clinic in order to extract the coordinates of the joints of the patient over time and predict the decision parameters. Finally, having the decision parameters we pre-classify gait disorder intensity of a patient and as the result make decisions on whether a patient needs treatment or not.\\
\section{Related Works}
The research about the use of machine learning methods in biological studies has been a subject of great attention in the recent years \cite{Andrea, Alex, Ehsan}. Mannini et al. \cite{Andrea} used a general probabilistic modeling approach for the classification of different pathological gaits. They used a Support Vector Machines (SVM) classifier and found 90.5$\%$ of subjects assigned to the right group of gait disorder classification. Alex et al. \cite{Ehsan} used the measured electrical signals on the surface of cerebral cortex in 16 locations within time interval of 10 minutes in order to identify potential seizure occurrence. They used a Fast Fourier Transform (FFT) algorithm to get power density in some frequency frames and used it as their input features. In their case, SVM with Radial Basis Function (RBF) Kernel gave the best results.
\section{Dataset and Features}\label{data}

\begin{figure}[H]
\begin{subfigure}{.33\textwidth}
\centering
\includegraphics[width=.5\linewidth]{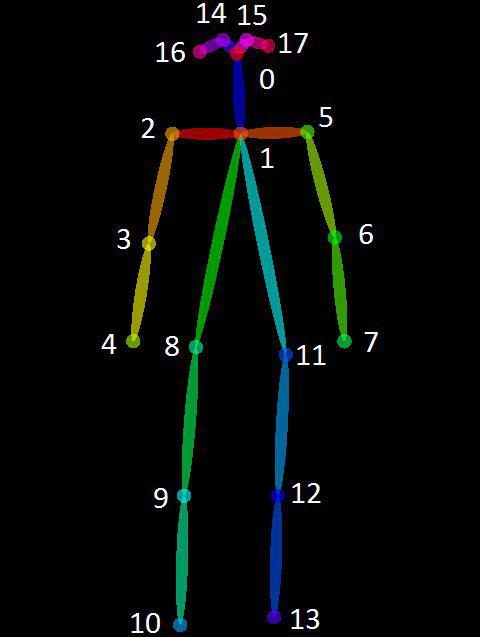}
\caption{}
\label{f1}
\end{subfigure}%
\begin{subfigure}{.33\textwidth}
\centering
\includegraphics[width=.95\linewidth]{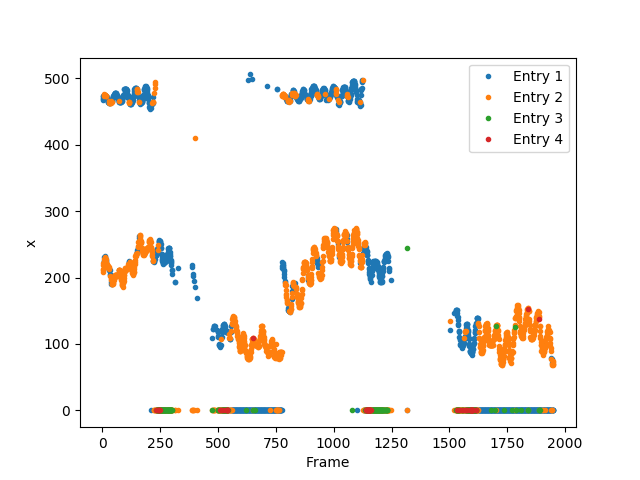}
\caption{}
\label{f2}
\end{subfigure}%
\begin{subfigure}{.33\textwidth}
\centering
\includegraphics[width=.95\linewidth]{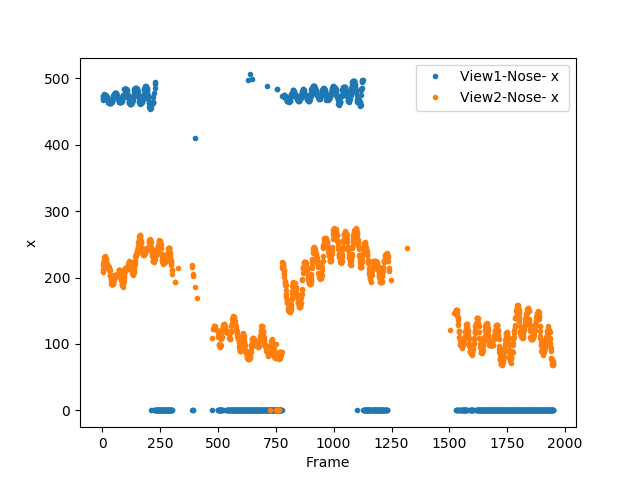}
\caption{}
\label{f3}
\end{subfigure}
\caption{(a) Numbering of joints of the body by OpenPose (b) $x$-coordinate of nose appeared in JSON files for 2000 frames (c) $x$-coordinate of nose from two camera views after cleaning}
\label{fig:fig}
\end{figure}

Source of the data used in this project come from Gillette Children's Specialty Health-care clinic. The data source is composed of two parts, in the first part we have diagnosis results and clinical parameters measured by the doctors in the clinic for every patient. By using the first part of the data set, the goal is to extract clinical features (e.g. cadence, step length, and walking speed). From this data, seven of the most relevant features for gait disorder recognition are used such as the angle between the front of the shin and the top of the foot  (\textit{AnkleDorsi}),  the angle made by the long axis of the foot from the heel to 2nd metatarsal and the line of progression of gait (\textit{FootProg}), the measured speed of walking, the step length, the distance between two successive placements of the same foot (\textit{strideT}), a person's weight divided by height squared (\textit{strideTbmi}), and the ratio of time we are in stance phase( when foot remains in contact with the ground) over the gait period (\textit{percentStance}). \\
The second part of data is extracted from 5000 videos where patients for gait disorder diagnosis walk in front of two cameras with lateral and front views. Using a code package called \textit{OpenPose} \cite{simon2017,wei2016} coordinates of 18 important joints of body (see Fig.\ref{f1}) are extracted for the duration of every patient's walking test and stored as a \textit{JSON} file. For doing this, in the processing of provided data the first step consists of cleaning and separating the data for lateral and front views. To define a perfect data sample, we would have two entries in each JSON file (i.e. in each frame of the video), one for the lateral view and one for the front view, each of these two entries has 18 set of data points for each joint and each set has three values corresponding to the coordinate values ($x$, $y$) and a confidence indicator ($c$) assigned by OpenPose. The challenging part is that many JSON files have either fewer or more than two entries and this can possibly happen due to three scenarios. \textbf{(I).} Having more than one person in the video which can happen when a second or third person appears in the video to accompany the patient during the walk. \textbf{(II).} Having shadows in the video which makes OpenPose to mistakenly consider the shadow of a person on the wall as a person. \textbf{(III).} Patient being out of one camera view which makes OpenPose not to record the positions of the joints of one of the views, since patient might be out of that camera view.\\

Fig.\ref{f2} highlights some of the above-mentioned issues where the $x$ coordinate of a patient's nose is plotted. It can be seen that there are up to 4 entries in this video, and each entry appears/disappears in different parts of the video. Also, we can see the flipping of positions between entry 1 and 2 for the video segment $[250, 500]$. Our strategy for data cleaning is to keep track of the patient's center of gravity in each view and for each new frame we judge the 4 entries based on their distance from this position. The center of gravity is calculated using only the joint points $ \lbrace 0, 1, 2, 5, 8, 11, 14, 15, 16, 17 \rbrace$ (cf. Fig.\ref{f1}), since these positions are the ones that move the least relative to the center of the body. Fig.\ref{f3} shows the results of our cleaning, we see that the data has been separated according to the left and right positions. We also see that the parts from different entries have been combined into our two final results. This cleaning may fail and its failure can be due to the absence of one view in the first few frames. Since right and left view contains information from the same patient we omitted left view and only used right view for data samples. In order to extract features, after having the cleaned time series of coordinates of a patient's body, in the first step we normalize (scale) the coordinates with respect to the distance between a patient's hip and neck. This helps to avoid misinterpretation of videos by discarding the influence of variations in the distance of different patients from camera while walking. Afterwards, using Fast Fourier Transform (FFT) we get the frequency spectrum of each joint. Next, from the FFT results for each right and left ankle we do weighted integration for 12 frequency intervals $(f_i,f_{i+1})$ as $a_i=\int_{f_i}^{f_{i+1}}A^2fdf$ and quantify the initial time series of the ankle joints in terms of power densities, where $f$ is the frequency, $A$ is the amplitude, and $a_i$ is the corresponding feature of each frequency interval. Finally, by doing this we come to 24 features (12 for each ankle) which will be used for predictions in section \ref{sec_vid}
\section{Methods and Results}
\subsection{Prediction Using Clinically Measured Data}
\subsubsection{Regression Models}
Different regression methods have been tried out mainly using MATLAB learning toolboxes. A summary of our results is presented in following sub-sections. To find the most useful features we experimented with various feature selection methods and chose the feature set showing promising results and mentioned in section \ref{data}. K-fold cross-validation with 5 folds was used to prevent over-fitting. It is worth mentioning that Principal Component Analysis (PCA) did not lead to any improvements in the results. We suspect the small number of features selected to be the cause, as PCA will further cut this number and degrades the performance since the resulting features set is not sufficient to predict the cadence with high accuracy.

\paragraph{Linear Regression (LR)}
This method attempts to minimize sum of squares of differences between the observed and predicted values by fitting the optimal linear predictor on the data. Among two different variations of linear methods used, step-wise regression achieved smaller mean squared error (MSE) than simple regression, since it uses forward selection to choose the predictive features that have the highest correlation with the target value as you see in table \ref{tab:linear}. This model also allowed us to quickly analyze the impact of the decisions we made in terms of pre-processing the data and creating feature vectors.

\begin{table}[h!]
\begin{center}
\resizebox{0.5\textwidth}{!} {
\begin{tabular}{*{5}{c}}
\hline
Model & RMSE & $R^2$ & MSE & MAE \\ [0.5ex]
\hline\hline
Linear Regression & 4.5\% & 85\% & 0.2\% & 2.9\%\\
\hline
Step-wise Linear Regression & 2.5\% & 95\% & 0.06\% & 1.3\%\\
\hline
Random Forest & 2.4\% & 96\% & 0.06\% & 1.08\%\\
\hline
Linear SVM & 4.9\% & 83\% & 0.24\% & 2.7\%\\
\hline
Medium Gaussian SVM & 4.38\% & 86\% & 0.19\% & 2.37\%\\
\hline
\end{tabular}
}
\end{center}
\caption{Linear regression, Random forest, and Support Vector Machine (SVM) results} \label{tab:linear}
\end{table}


\paragraph{Random Forest (RF)}
 Random forest (RF) is an ensemble learning method that trains a model by creating decision trees based on features and then reporting either the mode (for classification) or the average (for regression) of the results of all the decision trees in the model. The results of regression using Random Forest are in table \ref{tab:linear} where we are reporting the best performance that we got after trying different variations by changing the number and depth of decision trees. It can be seen that RF outperforms linear regression by a factor of 10 in MSE.


\paragraph{Support Vector Machine (SVM)}

In an attempt to improve the results obtained by the previous methods, support vector machine (SVM) using stochastic gradient descent (SGD) to minimize hinge loss was used. Linear SVM results shown in table \ref{tab:linear} for cadence predictions were comparable to those produced using linear regression shown in table \ref{tab:linear} where we tried to avoid overfitting slightly by reducing the training algorithm step size, adjusting the regularization parameter, and reducing the number of training iterations in our models. Gaussian SVM which uses a Gaussian kernel to produce curved boundaries was also used and found to outperform the linear SVM by about 17\% improvement in MSE (see table \ref{tab:linear}). This can be due to the fact that Gaussian SVM can capture well the inherent non-linear relationship between different features.


\subsubsection{Neural Networks (NN)}

\begin{figure}[!h]
\centering
\includegraphics[width=0.35\textwidth]{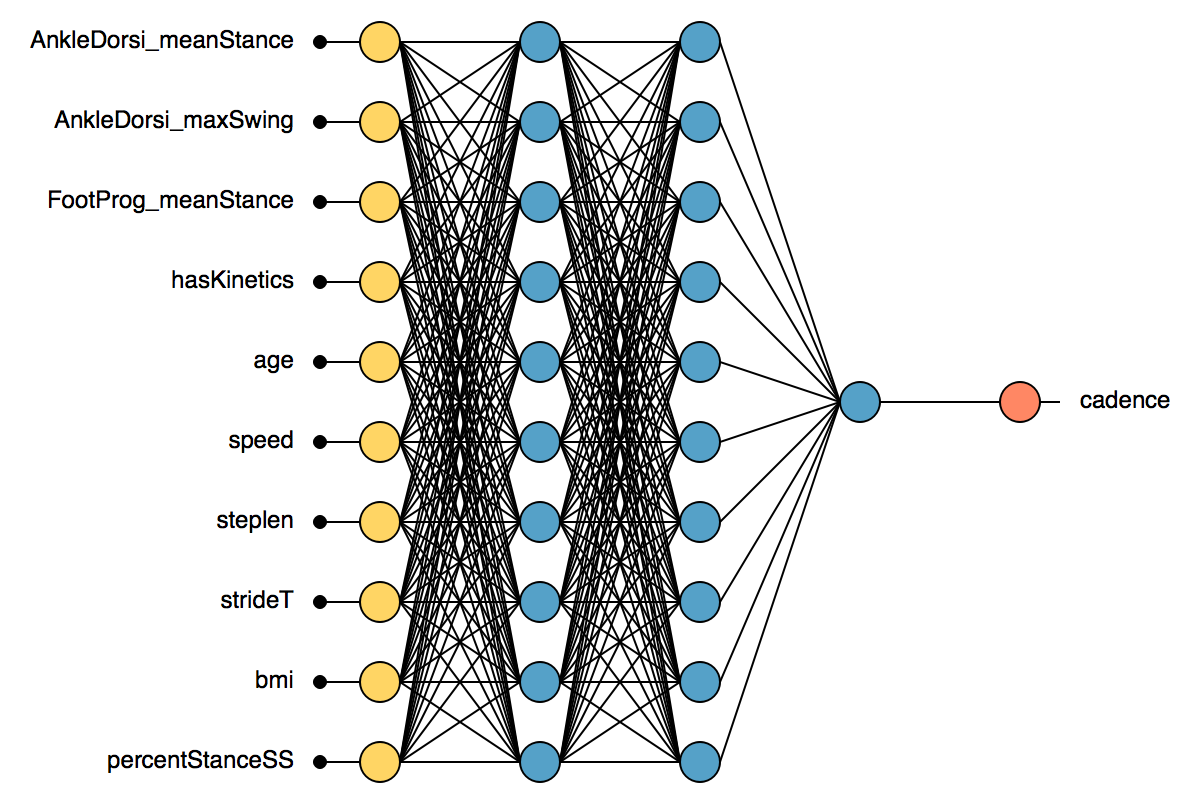}
\caption{Neural Network Architecture}
\label{NN}
\end{figure}

Neural networks (NN) despite their simplicity, are able to extract the relationship between different features and produce new ones that capture the correlation between data features and this way they outperform most of the current machine learning techniques. For this part, we used MATLAB's Neural Net Fitting simulator. Each feature vector was normalized by dividing the values by the maximum value for that specific feature. Then we shuffled the values and chose [70\%, 15\%, 15\%] metric for training, developing, and testing data samples respectively. For the neural network, we considered different architectures and chose the one which gave the lowest training MSE for fewer iterations while using K-fold cross-validation with 5 folds. The final architecture was a three-layered neural network shown in Fig. \ref{NN} with hyperbolic tangent function activations for all layers except the first, which was linear. The results for train, validation, and test data are shown in table \ref{tab:neural}.
\begin{table}[h!]
\begin{center}
\begin{tabular}{ *{3}{c} }
\hline
Data type & MSE & Regression\\
\hline \hline
Train & 0.0102\% & 99.62\% \\
\hline
Validation & 0.0098\% & 99.65\% \\
\hline
Test & 0.00186\% & 99.28\% \\
\hline
\end{tabular}
\end{center}
\caption{Neural networks results} \label{tab:neural}
\end{table}

%

\begin{figure}[!h]
\centering
\begin{subfigure}{.23\textwidth}
\centering
\includegraphics[width=1\linewidth]{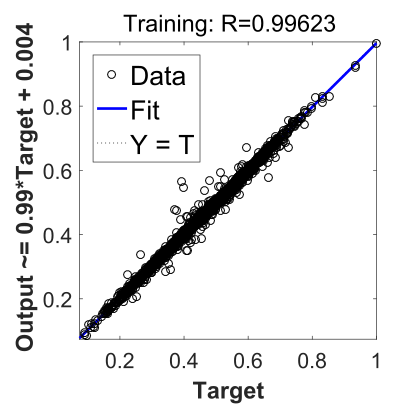}
\caption{Training Data}
\label{}
\end{subfigure}
\begin{subfigure}{.23\textwidth}
\centering
\includegraphics[width=1\linewidth]{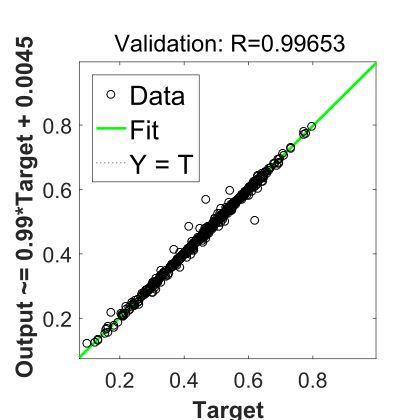}
\caption{Validation Data}
\label{}
\end{subfigure}
\begin{subfigure}{.23\textwidth}
\centering
\includegraphics[width=1\linewidth]{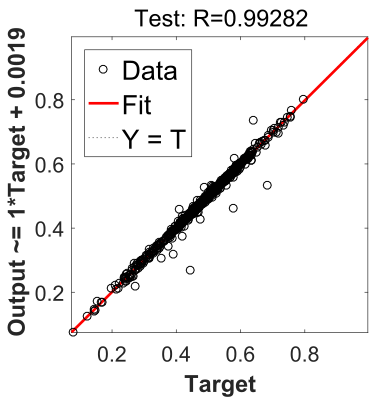}
\caption{Test Data}
\label{}
\end{subfigure}
\begin{subfigure}{.23\textwidth}
\centering
\includegraphics[width=1\linewidth]{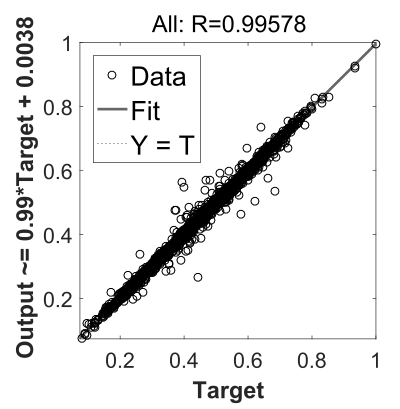}
\caption{All Data}
\label{}
\end{subfigure}
\caption{Prediction of cadence using Neural Network }
\label{fig:fig}
\end{figure}

%

\subsubsection{Classification Models}

Our final goal in this project is to be able to predict if treatments are needed for a patient or not. In our first group of data set (lab measured data) \textit{Gross Motor Function Classification (gmfcs)}. Gross motor skills are the abilities required in order to control the large muscles of the body for walking, running, sitting, crawling, and other activities. This classification system has 5 levels. The higher your level is the more difficulty you have in doing certain actions specified in gross motor skills. MATLAB Classification Learner toolbox was used for this task. We chose features using forward selection as predictive values and tried different variations of Random Forest, KNN, SVM, and an Ensemble of each one.  The highest resulting accuracy was achieved by Ensemble of boosted trees as 71.8\%; however, all the values in the two last classes are being misclassified. This might be due to a small number of data in those classes and possible mismatch between the training and testing sub-samples. So the classifier, clusters most of them into the first predicted class corresponding to no need for treatment. Since reaching conclusive results was hard based on the confusion matrix due to incoherency between accuracy of different clusters, we tried clustering different level of \textit{gmfcs} together. Specifically, we kept first class intact, but clustered the second and third group in one bin indicating low level of gait disorder and third, fourth, and fifth group into another bin indicating high level of gait disorder (i.e. needing treatment). The results show improvement in accuracy with 8.2\% increase. In this case the Ensemble of RUSboosted trees gave us the best result. RUSBoost is an algorithm that handles class imbalance problem in data with discrete class labels. It uses a combination of RUS (random under-sampling) and the standard boosting procedure, to better model the minority class by removing majority class samples. 



\begin{table}[h!]

\begin{center}

\resizebox{0.7\textwidth}{!} {

\begin{tabular}{*{4}{c}}

\hline

Model & Accuracy & Area under curve (AUC) & Correctly classified\\ [0.5ex]

\hline \hline

Coarse Tree & 75.5\% & 86\% & 50.29\%\\

\hline

Linear Discriminant & 57.6\% & 76\%  & 48.17\% \\

\hline

Linear SVM & 58.7\% & 55\% & 391.0\% \\

\hline

RUSBoosted Trees (Ensemble) & 80\% & 92\% & 53.39\% \\

\hline

\end{tabular}

}

\end{center}

\caption{Treatment classification} \label{tab:classify}

\end{table}

%

%

\begin{figure}[!h]
\begin{subfigure}{.5\textwidth}
\centering
\includegraphics[width=0.65\linewidth]{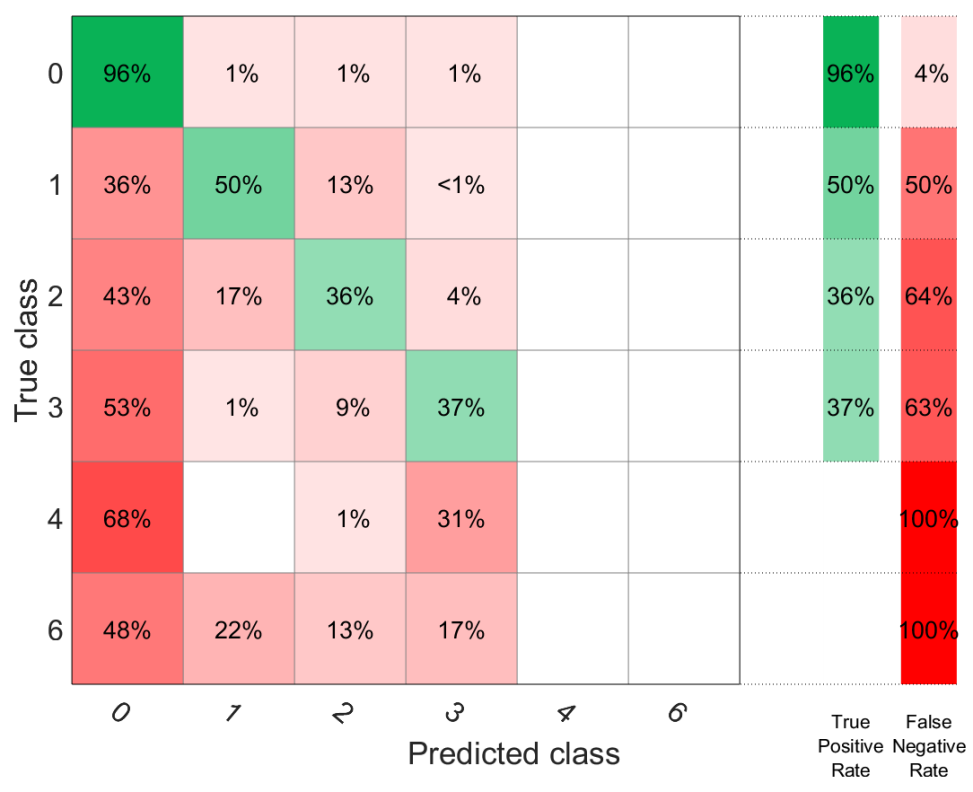}
\caption{Unclustered}
\label{}
\end{subfigure}%
\begin{subfigure}{.5\textwidth}
\centering
\includegraphics[width=0.65\linewidth]{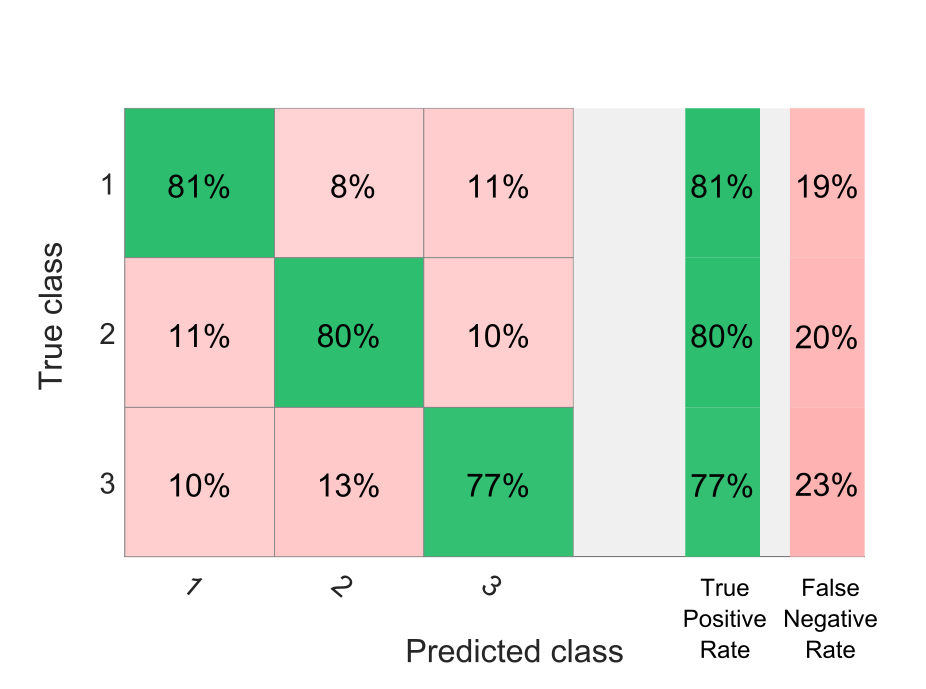}
\caption{Clustered}
\label{}
\end{subfigure}
\caption{Confusion matrix for gait disorder recognition}
\label{fig:fig}
\end{figure}
\subsection{Prediction Using Video Data}\label{sec_vid}
In this part, the data extracted from videos will be used to predict decision parameters for gait disorder recognition . Aside from the fact that the features for this part are obtained using FFT and weighted integration (cf. Section \ref{data}), two other differences exist between this part and the previous one (i.e. clinically measured data). Firstly, here we predict three decision parameters such as cadence, step length, and speed and secondly, the tools used here are  Python libraries scikit-learn and Keras (NN) instead of MATLAB.
\subsubsection{Linear Regression (LR)}
Regularized linear regression was used as our first model to be trained on the collected data (shuffled). The features where first normalized and the complexity of our features' space suggests to perform a principal component analysis (PCA) on our features by projecting them on a reduced space which gives us an uncorrelated set of features. Doing this reduced our features from 26 to 6. The reduced set of features was used to predict cadence, step length, and speed. Mean square error and the coefficient of determination($R^2$) metrics were used to evaluate our trained models, the evaluation is done on a set of test data that has not been seen by the model. We experimented with different values for the regularization parameter to see whether increasing regularization would lead to any improvements and our conclusions are that it has no impact. The coefficient of determination results for the three targets we wanted to predict (cadence, step length, and speed) show a very low correlation, the cadence performs slightly better ($R^2=0.14$ for test set) while predicting the speed and the step length remains worse than simply taking the average. The results for the linear regression were expected as our features non-linearity and complexity cannot be captured by simple linear regression. The cadence gives better results because of the design of the features which is based on the frequency (cadence is the frequency of steps).
\begin{table}[h!]
    
    \begin{center}
    
    \begin{tabular}{ *{3}{c} }
    
    \hline
    
    Targets & MSE & $R^2$ \\
    
    \hline \hline
    
    Cadence & 0.07 & 0.14\\
    
    \hline
    
    Step Length & 0.018& -0.2\\
    
    \hline
    Speed & 0.14 & -0.6 \\
    \hline
    
    \end{tabular}
    
    \end{center}
    
    \caption{Linear regression results for test set} \label{tab:neural}
    
\end{table}
\subsubsection{Support Vector Machine (SVM)}
The SVM for regression model was used with a radial basis function (RBF) kernel, we also experimented with linear and polynomial kernels and RBF gives the best results. 
We normalized the data in this case, and used PCA to reduce the parameters’ space.  
Regarding the hyperparameters, we optimized based on the two parameters $C$, 
which characterizes the penalty related to points far from the regression curve and $\gamma$ which is a kernel parameter used to configure the sensitivity 
to the norm of the feature vectors. We used the coefficient of determination ($R^2$) metric to evaluate the different models. 
Table \ref{tab1} gives the best $C$ and $\gamma$ parameters for the three targets: cadence, step length, and speed. 
Table \ref{tab2} gives the values for the mean square error and the coefficient of determination metrics for the three targets using the optimized hyperparameters, these results are for the test data set. The results are much better than the linear regression, 
we can see that for the speed prediction, the coefficient of determination is around $50\%$. The improvement for the prediction was expected, 
but it is surprising that the results for speed prediction are better than the cadence, the reason is that the way of extraction of the features using FFT is expected to be more suitable for cadence prediction due to its intrinsic frequency nature.
\begin{table}[h!]

   \begin{subtable}[t]{0.48\textwidth}
      \begin{center}
 
    \begin{tabular}{ *{3}{c} }
    
    \hline
    
    Targets & C& $\gamma$ \\
    
    \hline \hline
    
    Cadence & 1& 1\\
    
    \hline
    
    Step Length & 0.1 & 1\\
    
    \hline
    Speed & 1 & 1 \\
    \hline
    
    \end{tabular}
    
    \end{center}
    
    \caption{Optimal Hyperparameters for SVM} \label{tab1}
   \end{subtable}
   \begin{subtable}[t]{0.48\textwidth}    
       \begin{center}
    \begin{tabular}{ *{3}{c} }
    
    \hline
    
    Targets & MSE & $R^2$ \\
    
    \hline \hline
    
    Cadence & 0.02 & 0.36\\
    
    \hline
    
    Step Length & 0.01& 0.31\\
    
    \hline
    Speed & 0.04 &0.5\\
    \hline
    
    \end{tabular}
    
    \end{center}
    
    \caption{SVM results for test set} \label{tab2}
  \end{subtable}    
  \caption{SVM test results} \label{tab3}
\end{table}
\subsubsection{Neural Network (NN)}
\begin{figure}[H]

\begin{subfigure}{.33\textwidth}

\centering
\includegraphics[width=0.9\linewidth]{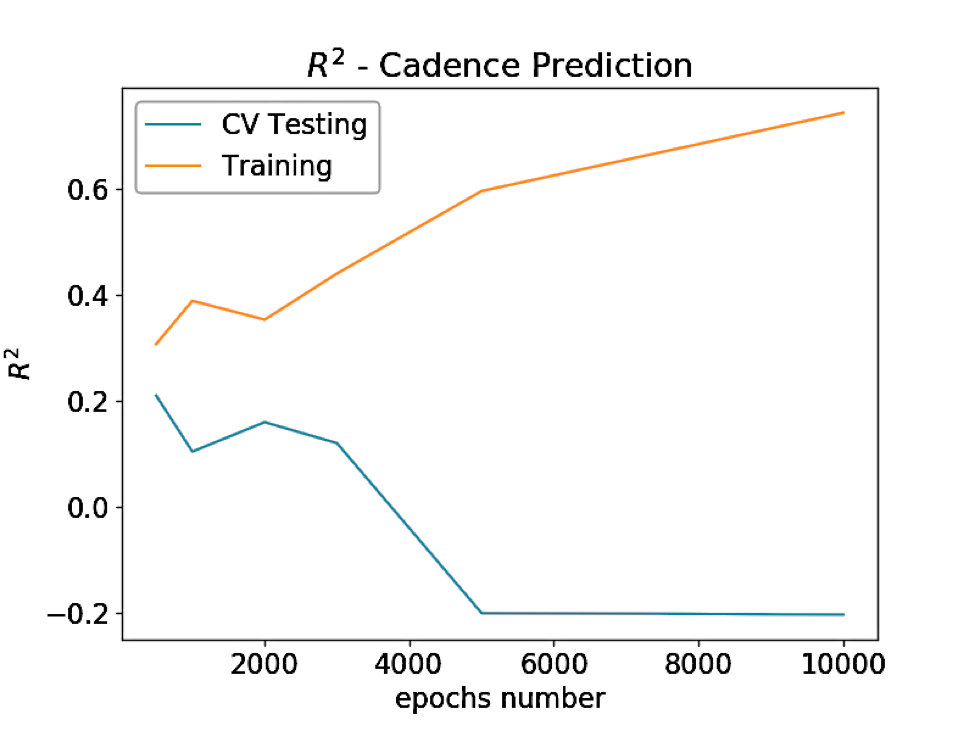}
\caption{}
\label{f2}
\end{subfigure}%
\begin{subfigure}{.33\textwidth}
\centering
\includegraphics[width=0.9\linewidth]{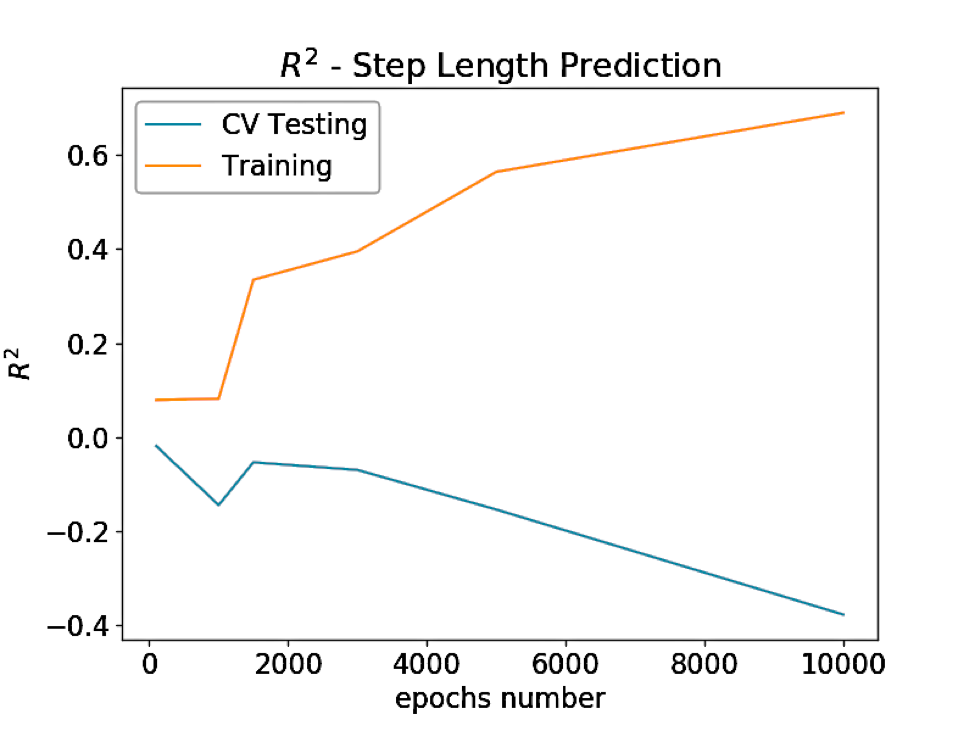}
\caption{}
\label{f3}
\end{subfigure}
\begin{subfigure}{.33\textwidth}
\centering
\includegraphics[width=0.9\linewidth]{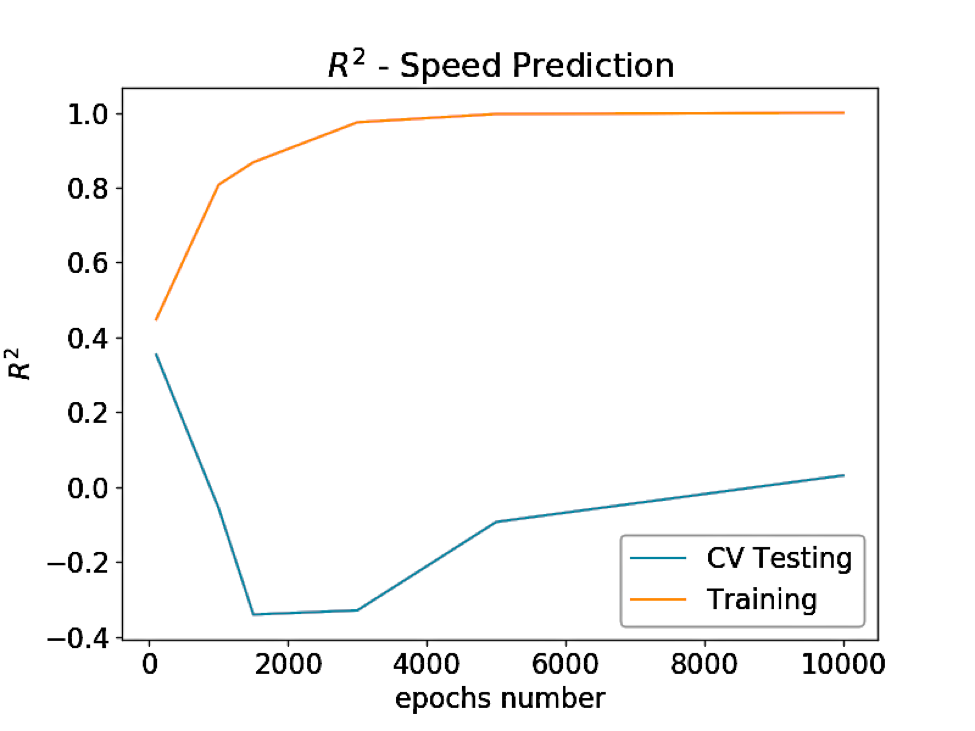}
\caption{}
\label{f3}
\end{subfigure}
\caption{NN test data $R^2$ metric for  (a) Cadence, (b) Step Length, and (c) Walking Speed}
\label{fig1}
\end{figure}

Neural network (NN) was the last model to be tested, while experimenting with the NN we only used the cadence target for parameters' choice, we will see that this had an impact on the prediction of the other two targets: speed and step length.  We experimented with a different number of layers, number of neurons per layer and activation functions, our final model has 3 hidden layers with respectively 50, 40 and 30 neurons per layers, and the activation function is a tangent hyperbolic. Hold out cross-validation was used for the training, the data was divided into 4 buckets, for each loop we train on 3 buckets and we test on the last one, then an average over the four different cases was calculated. Mean square error and the coefficient of determination ($R^2$) metrics were used to evaluate our trained models. We trained our model with a different number of epochs. In order for the simulation to reach a stabilized mean error (or to identify the moment of stopping) we used relatively high number of epochs, we will see that we have for all cases overfitting after an optimal number of epochs, the test errors we are reporting are at this optimal position.
The results for the test data are presented in Fig\ref{fig1} for the three target vectors: cadence, step length, and speed.  For the cadence, the optimal epoch number is 2000 which gives us a coefficient of determination ($R^2$) of $18\%$. For the speed and step length we have a serious problem of overfitting, this may be due to the fact that the neural network configuration (number of layers, number of neurons…) was not designed with these targets in mind (we used cadence).

\section{Conclusion and Future Work}
In this project we explored two approaches to predict main decision parameters in gait disorder recognition (i.e. cadence, step length, and speed) that can be used afterwards to give advice on whether the patient needs treatment or not. Due to the importance of cadence, we used the two approaches to predict it, while for the case of speed and step length we only made predictions through the video data. The predictions using the clinically measured data outperforms the ones using the video-extracted data. To improve the predictions using the video-extracted data, we should address the issues highlighted in section \ref{data} about data cleaning and also explore other techniques of feature extraction from time series.
\section{Acknowledgement}
At the end, the authors would like to acknowledge Dr. Lukasz Kidziski from Mobilize center at Stanford for facilitating our access to the dataset.
\bibliographystyle{unsrt}

\bibliography{references}

\begin{thebibliography}{1}

\bibitem{Andrea}
Andrea Mannini and Diana Trojaniello.
\newblock A machine learning framework for gait classification using inertial
  sensors: Application to elderly, post-stroke and huntington’s disease
  patients.
\newblock 2016.

\bibitem{Alex}
Alex Fu, Spencer Gibbs, and Yuqi Liu.
\newblock Seizure prediction from intracranial eeg recordings.
\newblock 2014.

\bibitem{Ehsan}
Ehsan Dadgar-Kiani.
\newblock Applying machine learning for human seizure prediction.
\newblock 2016.

\bibitem{simon2017}
Tomas Simon, Hanbyul Joo, and Iain Matthews.
\newblock Hand keypoint detection in single images using multiview
  bootstrapping.
\newblock 2017.

\bibitem{wei2016}
Shih-En Wei, Varun Ramakrishna, and Takeo Kanade.
\newblock Convolutional pose machines.
\newblock 2016.

\end{thebibliography}

\end{document}